\documentclass[pre,twocolumn,longbibliography]{revtex4-1}
\usepackage{enumitem,kantlipsum}
\usepackage{graphicx}
\usepackage{amssymb,amsmath,amsfonts}
\usepackage{epsf}
\usepackage{wrapfig}
\usepackage[normalem]{ulem}

\newcommand{\R}{\mathbb{R}}

\renewcommand{\matrix}[2]{ \left(\begin{array}{#1} #2 \end{array}\right)}
  
\newcommand{\bra}{\langle}
\newcommand{\ket}{\rangle}
\newcommand{\mbx}{\mathbf{x}}
\newcommand{\mbX}{\mathbf{X}}
\newcommand{\LL}{\mathcal{L}}
\def\given{\:|\:}
\usepackage{color}

\newcommand{\bi}[1]{Fig.~\ref{fig:#1}}
\newcommand{\e}[1]{eq.~(\ref{eq:#1})}
\newcommand{\be}{\begin{equation}} 
\newcommand{\ee}{\end{equation}}

\usepackage{graphicx}
\usepackage{dcolumn}
\usepackage{bm}
\usepackage{color}
\usepackage{bbm} 
\newcommand{\lr}[1]{\left\langle #1 \right\rangle}
\newcommand{\se}[1]{sec.~\ref{sec:#1}}
\newcommand{\sse}[1]{subsec.~\ref{ssec:#1}}

\newcommand{\ba}{\begin{eqnarray}} \newcommand{\ea}{\end{eqnarray}}

\renewcommand{\vec}[1]{\mathbf{#1}}

\begin{document}
\title{Phase descriptions of a multidimensional Ornstein-Uhlenbeck process}
\author{Peter J.~Thomas}
\affiliation{%
Department of Mathematics, Applied Mathematics, and Statistics.\\
Case Western Reserve University,
Cleveland, Ohio, 44106, USA
}%
\author{Benjamin Lindner}
\affiliation{Bernstein Center for Computational Neuroscience and Department of Physics. \\
Humboldt University, 10115 Berlin, Germany.}
\date{\today}                                           
\begin{abstract}
Stochastic oscillators play a prominent role in different fields of science. Their simplified description in terms of a phase has been advocated by different authors using distinct phase definitions in the stochastic case. One notion of phase that we put forward previously, the \emph{asymptotic phase of a stochastic oscillator}, is based on the eigenfunction expansion of its probability density. More specifically, it is given by the complex argument of the eigenfunction of the backward operator corresponding to the least negative eigenvalue. Formally, besides the `backward' phase, one can also define the `forward' phase as the complex argument of the eigenfunction of the forward Kolomogorov operator corresponding to the least negative eigenvalue. Until now, the intuition about these phase descriptions has been limited. Here we study these definitions for a process that is analytically tractable, the two-dimensional Ornstein-Uhlenbeck process with complex eigenvalues.
For this process, (i) we give explicit expressions for the two phases; (ii) we demonstrate that the isochrons are always the spokes of a wheel, but that (iii) the spacing of these isochrons (their angular density) is different for backward and forward phases; (iv) we show that the isochrons of the backward phase are completely determined by the deterministic part of the vector field, whereas the forward phase also depends on the noise matrix; and (v) we  demonstrate that the mean progression of the backward phase in time is always uniform, whereas this is not true for the forward phase except in the rotationally symmetric case. We illustrate our analytical results for a number of qualitatively different cases.
\end{abstract}
\maketitle

\section{Introduction}

Both oscillations and noise are ubiquitous in many systems of interest in physics \cite{HemLax67,Str67}, biology \cite{MarBoz03,KumKra05,EloLei00} and neuroscience \cite{Ermentrout2014Encyc,LaingLord2010}.  Synchronization and entrainment of deterministic oscillators may be analyzed by finding a coordinate transformation to a one-dimensional phase variable defined in terms of the asymptotic phase of a limit cycle \cite{BrownMoehlisHolmes2004NeComp,Guckenheimer1975JMathBiol,Kuramoto1975chapter,Schwemmer11}.  The classical approach breaks down in several  cases of interest including: (a) limit-cycle systems endowed with noise \cite{HemLax67,EbeHer86,Lin02,UshWue05,GleDoh06,JulDie09,StiefelFellousThomasSejnowski2010EJNsci,DitlevsenGreenwood2013JMB},  (b) spiral-sink systems with noise-sustained oscillations, also known as quasicycles \cite{UhlOrn30,LSGZue90,LugMcK08,WalBen11,BroBre15}, and (c) systems with an attracting heteroclinic cycle \cite{ShawParkChielThomas2012SIADS,ThoLin14,BalTho17}.

Consider as an example for (b), the two-dimensional Ornstein-Uhlenbeck process in the form of two linear stochastic differential equations
\ba
\label{eq:example1}
\frac{dx_1}{dt}&=A_{11} x_1 + A_{12} x_2 +B_{11} \xi_1  +B_{12}\xi_2\nonumber\\
\frac{dx_2}{dt}&=A_{21} x_1 + A_{22} x_2  +B_{21} \xi_1 +   B_{22}\xi_2.
\ea
In these equations the $\xi_i$ represent independent delta-correlated Gaussian white noise sources with $\langle\xi_i(s)^T\xi_j(t)\rangle=\delta_{ij}\delta(s-t)$.
A trajectory  for a particular choice of parameters, corresponding to a spiral sink, is shown in \bi{example1}.   This system resembles a well behaved oscillator in many respects: it displays a noisy rotation around the origin in phase space (A), shows roughly oscillatory behavior also in the single components (B), and exhibits a pronounced peak at a non-vanishing frequency in the power spectrum of one of the components (C). The quality factor of the oscillation (frequency of the spectral peak divided by its width at half maximum height), is rather high in the chosen example ($Q_\text{F}\approx 10$). Despite the apparent oscillations, however, considering the noiseless system (setting $B_{ij}=0$ in \e{example1}), we cannot define a phase of this spiral-sink system (see below \sse{deterministic_setting}).
   
\begin{figure*} 
   \centering
   \includegraphics[width=0.7\textwidth]{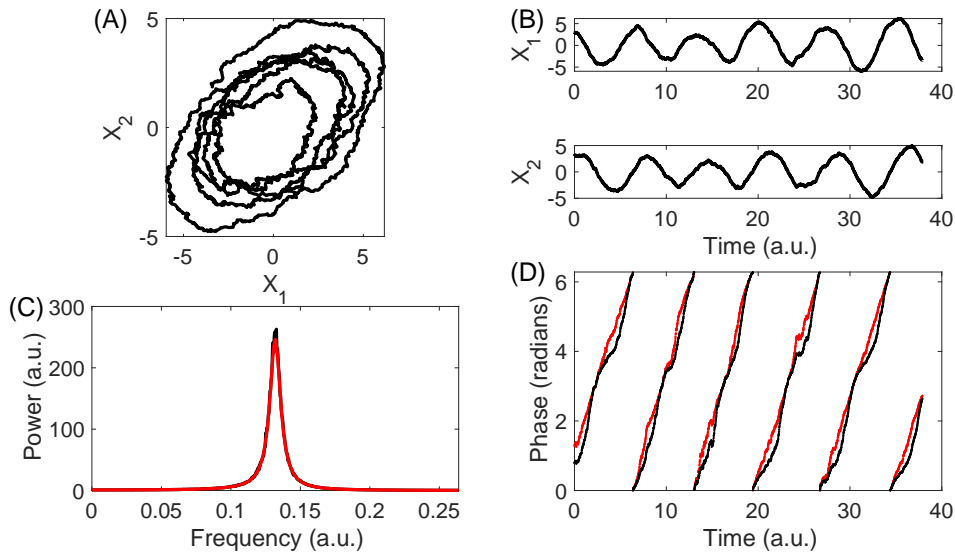} 
   \caption{\textbf{An example for the 2d Ornstein-Uhlenbeck process.} 
We simulated \e{example1}  with the parameters $A_{11}=0.47, A_{12}=-1.25, A_{21}=0.75, A_{22}=-0.53$ and the noise values $B_{11}=0.4861,B_{12}=B_{21}=-0.1169,B_{22}=0.3692$;
all shown values of $x_1$ and $x_2$ in arbitrary units. Trajectory in phase space $(x_1,x_2)$ (A), single trajectories as time series (B). We used a simple stochastic Euler procedure for numerical integration.
Power spectrum (C) of $x_1(t)$ ($400$ realizations of time series  of $2^{20}$ steps with a time step of $\Delta t=2.5\cdot 10^{-3}$; analytical result, \e{power_spectrum}, is shown by a solid line.       
The asymptotic phase  of the system, $\Psi(t)=\Psi(x_1(t),x_2(t))$ [red line, see \e{BW_phase}] is compared to the geometric phase $\vartheta(t)=\arctan(x_2(t)/x_1(t))$ [black line];  both are shown as functions of time (D). 
   }
   \label{fig:example1}
\end{figure*}

In \cite{ThoLin14} we introduced a new definition of the asymptotic phase for robustly oscillatory stochastic systems, based on a spectral decomposition of the backward (or adjoint) Kolmogorov density operator. We note that an equivalent decomposition was independently introduced in the context of dephasing of genetic oscillators in \cite{PotoyanWolynes2014PNAS} and an alternative definition of the phase for stochastic oscillators, based on a  mean first passage time construction, was put forward in \cite{SchwabedalPikovsky2013PRL}; 
in this paper we exclusively focus on the asymptotic phases based on the eigenfunction expansion.

Our asymptotic phase is well defined for noisy systems whether the underlying mean-field dynamics exhibits a stable limit cycle, spiral sink or a stable heteroclinic orbit, provided the eigenvalues and eigenfunctions of the Kolmogorov operator and its adjoint operator satisfy a set of natural conditions \cite{ThoLin14}, detailed below.  Moreover, in the case of a stable limit cycle system, when the classical asymptotic phase is well defined, the isochrons for the stochastic system with small noise levels closely resemble the isochrons of the deterministic system.

Many questions regarding the asymptotic phase and  the forward phase are still open. 
We lack intuition about the shape of the isochrons, their dependence on the noise strength, and the difference between the backward and forward phases. The purpose
of this paper is to explore these issues for a case where we can obtain analytical insights, that is, the two-dimensional Ornstein-Uhlenbeck process with complex eigenvalues of the drift matrix (a special case of \e{example1}). For this system, we derive explicit expressions for the asymptotic (backward) phase and for the forward phase.  We hope to convince the reader that this example is nontrivial. Inter alia, we find qualitative differences in the forward and backward phases. For instance, the backward phase isochrons do not depend at all on the noise properties of the system, i.e.~on $B_{ij}$, whereas the forward phase's isochrons do. Furthermore, if we plot the asymptotic phase $\Psi(t)=\Psi(x_1(t),x_2(t))$ as a function of time (see \bi{example1}D), it progresses  more steadily than the geometric phase $\vartheta(t)=\arctan(x_2(t)/x_1(t))$. In fact, we will show that the \emph{average} rate of increase for the backward phase does not depend on time, nor on the geometric phase.

Our paper is organized as follows. In the next section we recall the definitions of the backward and forward phases for stochastic systems, as well as the definition of the deterministic phase. In \se{model} we introduce the two-dimensional Ornstein-Uhlenbeck model and our specific parametrization of it. In \se{analytic} we derive the two eigenfunctions of interest, extract the dependence of backward and forward phases on the geometric phase, and discuss their general properties. In \se{illustration} we look at a number of examples that give  some insight into  the differences between the two phase types. We conclude in \se{conclusions} with a brief discussion and outlook.  

Table \ref{tab:notation} provides a list of notation.

\begin{table}[htbp]
   \centering
   \begin{tabular}{ll} 
   Symbol & Meaning \\ \hline
   $\tau\in[0,T)$ & Asymptotic phase of deterministic LC (timelike).\\
   $\theta\in[0,2\pi)$ & Asymptotic phase of deterministic LC (circular).\\
   $\vartheta\in[0,2\pi)$ & Geometric phase (standard polar coordinates).\\
   $\Psi\in[0,2\pi)$ & Backward phase for stochastic oscillator. \\ 
   $\Phi\in[0,2\pi)$ & Forward phase for stochastic oscillator.\\ \hline
   \end{tabular}
   \caption{Notation for different phase notions used in the paper.  Limit cycle period is $T$. }
   \label{tab:notation}
\end{table}

\section{Review of the phase definitions in deterministic and stochastic systems}
\label{sec:phase_definition}

In this section we recall notions of the asymptotic phase of an oscillator for deterministic and stochastic dynamical systems.  For an $n$-dimensional dynamical system, the reduction to a $1$-dimensional phase description assigns a scalar variable (the phase) to each point in the underlying space.  For a deterministic limit cycle oscillator, the phase variable $\theta$ should satisfy $d\theta/dt=\text{const}$ along trajectories.    In the setting of stochastic dynamical systems, we can obtain a reduced description of an oscillatory Markov process in terms of the eigenfunctions of the generator of the process.  

\subsection{Stochastic setting}
\label{ssec:stochastic_setting}
%
As in \cite{ThoLin14}, we assume the state of the system is given by a point $\mbX$ in a (discrete or continuous) finite dimensional space, $\mbX\in\mathcal{X}$, and evolves in continuous time $t\in\R$ as a Markov process with transition density
\begin{equation}
P(\mbx,t\given \mbx_0,t_0)=\frac{1}{|d\mbx|}\Pr\{\mbX(t)\in[\mbx,\mbx+d\mbx) \given \mbX(t_0)=\mbx_0\}
\end{equation}
where $t>t_0$. We consider  processes that are homogeneous in time, meaning $P(\mbx,t\given \mbx_0,t_0)=P(\mbx,t-t_0\given \mbx_0,0)$.

We assume the density evolves according to a differential operator $\LL$ with formal adjoint $\LL^\dagger$.  That is, the evolution with respect to the latter time $t$ is given by a forward Kolmogorov equation (or forward Fokker-Planck equation, in the case of a system driven by additive Gaussian white noise)
\begin{equation}
\frac{\partial }{\partial t}P(\mbx,t\given \mbx_0,t_0) =\LL_{\mbx}[P(\mbx,t\given \mbx_0,t_0)]
\end{equation}
while the evolution  with respect to the earlier time $t_0$ is given by a backward or adjoint Kolmogorov equation (or backward Fokker-Planck equation, in the Gaussian case), 
\begin{equation}
-\frac{\partial }{\partial t_0}P(\mbx,t\given \mbx_0,t_0) =\LL^\dagger_{\mbx_0}[P(\mbx,t\given \mbx_0,t_0)].
\end{equation}
The operator $\LL^\dagger$ is also known as the \emph{generator} of the Markov process 
\cite{EthierKurtz2009book}.
We will assume the forward and backward operators $\LL$ and $\LL^\dagger$ have a  biorthogonal set of eigenfunctions $P_\lambda$, $Q^*_\lambda$ satisfying
\begin{eqnarray}
\LL[P_\lambda]&=&\lambda P_\lambda\\
\LL^\dagger[Q^*_\lambda]&=&\lambda Q^*_\lambda\\
\bra Q_\lambda | P_{\lambda'}\ket &=&\int_\mathcal{X} Q^*_\lambda(\mbx)P_{\lambda'}(\mbx)\,d\mbx=\delta_{\lambda,\lambda'},
\label{eq:biorthog}
\end{eqnarray}
where $\delta_{\lambda,\lambda'}$ is the Kronecker delta. 
We consider the case in which this biorthogonal system is complete, in the sense that for any $\mbx\in\mathcal{X}$, the conditional density may be written in terms of the spectral decomposition
\begin{equation}
P(\mbx,t\given \mbx_0,t_0)=P_0(\mbx)+\sum_\lambda e^{\lambda(t-t_0)}P_\lambda(\mbx)Q^*_\lambda(\mbx_0).
\end{equation}
We assume the system has a unique invariant probability distribution corresponding to the forward eigenfunction $P_0$ of the trivial eigenvalue $\lambda_0=0$; the corresponding backward eigenfunction is $Q_0\equiv 1$.  
We moreover assume the remaining eigenvalues, which may be real or complex, have negative real part.  

As in  \cite{ThoLin14}, we will define a system to be ``robustly oscillatory" if, in addition, it fulfills the following conditions: 
\begin{enumerate}[leftmargin=*,labelindent=0pt,itemsep=0pt]
\item The eigenvalue with  least negative real part forms a complex conjugate pair $(\lambda,\lambda^*)$, with $\lambda=\mu+i\omega$ and $\omega>0$.
\item The pair $(\lambda,\lambda^*)$ has real part  sufficiently far from the rest of the eigenvalue spectrum, specifically  $\Re\lambda'\le 2\mu$ for all $\lambda'\not=\mu\pm i\omega$. 
\item The relaxation rate of the probability density is significantly slower than the oscillation, in the sense that $|\mu|\ll \omega$.
\end{enumerate}
Under these assumptions, we may write the eigenfunction pair corresponding to the slowest decaying eigenvalue, $\lambda$, in complex notation as 
\be
P_1(\mbx) = v(\mbx) e^{-i\Phi(\mbx)}, \;\;\; Q^*_1(\mbx_0) = u(\mbx_0) e^{i\Psi(\mbx_0)},
\ee
which can be regarded as the definitions of the forward ($\Phi$) and backward ($\Psi$) asymptotic phases of the system.  That is, we may define the backward and forward phases for a point  $\mbx$ as functions of the coordinates $\mbx$ as 
\be
\Psi(\mbx)=\arg\left[ Q_1^*(\mbx)\right],\quad\Phi(\mbx)=-\arg\left[ P_1(\mbx)\right],
\ee
where $\arg(a+ib)=\tan^{-1}(b/a)$ is the complex argument of $z=a+ib$.   
As shown in \cite{ThoLin14}, the long-term behavior of the probability density, as it approaches the steady-state distribution $P_0(\mbx)$, will be dominated by the difference between the backward phase at the initial point $\Psi(\mbx_0)$   and the forward phase at the later point $\Phi(\mbx)$:
\begin{align}
\label{eq:P_asymp}
&\frac{P(\mbx,t\given \mbx_0,t_0)-P_0(\mbx)}{2u(\mbx_0)v(\mbx)}\simeq \\
&\quad e^{\mu(t-t_0)}\cos\left[\omega(t-t_0)+\Psi(\mbx_0)-\Phi(\mbx) \right],\text{ as }t\to\infty.\nonumber
\end{align}
This asymptotic behavior allows one to extract the forward and backward phase from numerical simulations of the stochastic process, and in principle also from data.

When considering a stochastic trajectory $\mbX(t)$ we may alternatively represent the phases as  functions of time
\be
\Psi(t)\equiv\Psi(\mbX(t)),\quad\Phi(t)\equiv\Phi(\mbX(t)),
\ee 
i.e.~as stochastic processes in their own right.  

\newcommand{\mbF}{\mathbf{F}}

\subsection{Deterministic setting}
\label{ssec:deterministic_setting}

Here, instead of an evolving density $P(\mbx,t\given\mbx_0,t_0)$, we consider a trajectory $\mbx(t)$ satisfying an ordinary differential equation with initial condition $\mbx_0$:
\be
\frac{d\mbx}{dt}=\mbF(\mbx),\quad\mbx(0)=\mbx_0.
\ee
If the system has a periodic solution with period $T$, that is, $\mbx(t)=\mbx(t+T)$, then  we may define a time-like phase $\tau$ as a map from the orbit $\Gamma=\{\mbx(t)\given 0\le t< T\}$ to the circle $\tau\in[0,T)$ satisfying the scalar differential equation 
\be
\label{eq:dtheta_dt}
\frac{d\tau}{dt}=1,\ee 
with initial condition  $\tau(0)=\tau_0$ \cite{ErmentroutTerman2010book}.  The phase $\tau_0$ associated with a reference point $\gamma_0\in\Gamma$ may be chosen arbitrarily, and $\tau$ is interpreted mod $T$.  
If all trajectories with initial conditions close to the periodic orbit converge to $\Gamma$ then we have a stable limit cycle $\mbx=\gamma(t)$. The set of initial conditions converging to $\Gamma$ is its \emph{basin of attraction}, $\mathcal{B}$.
For any initial condition  $\mbx_0\in\mathcal{B}$,
one defines its \emph{asymptotic phase}  as the scalar $\tau(\mbx_0)$ such that 
\be
\label{eq:asymptotic_phase}
\lim_{t\to\infty}\left|\mbx(t)-\gamma(t+\tau(\mbx_0)) \right| = 0,\text{ as }t\to\infty,
\ee
for the trajectory $\mbx(t)$ with initial condition $\mbx(0)=\mbx_0$. This phase reduction,  introduced in \cite{Guckenheimer1975JMathBiol}, has proven invaluable in the study of weakly coupled and weakly driven oscillators \cite{Kuramoto1975chapter,ErmentroutKopell1984SIAMJMathAnal,BrownMoehlisHolmes2004NeComp,HopIzh97,Schwemmer11}.
As originally discussed by Guckenheimer \cite{Guckenheimer1975JMathBiol}, the asymptotic phase may  equivalently be  described in terms of \emph{isochrons}, which are the level sets of a differentiable function $\mathcal{T}(\mbx)$ satisfying
\be
\label{eq:dTdt}
\frac{d\mathcal{T}(\mbx(t))}{dt} = \nabla\mathcal{T}(\mbx) \cdot\mbF(\mbx)=1,
\ee
for all points $\mbx\in\mathcal{B}$.  Under the natural boundary conditions imposed by continuity of $\mathcal{T}$ at the limit cycle, we may  identify $\tau=\mathcal{T}$, up to an additive constant (compare \e{dtheta_dt}).
For a limit cycle with period $T$, the normalization of the phase variable to the interval $[0,T)$ is conventional.  In order to emphasize the phase as a map to the circle, the phase normalization $\theta\equiv\left(2\pi \tau/T\right)\in[0,2\pi)$ may be used instead.  In this case \e{dTdt} is changed to $d\theta/dt=2\pi/T$.

Transient oscillatory activity may arise in deterministic models that do not possess limit cycles, for example spiral sink systems and stable heteroclinic cycles.  A deterministic dynamical systems has a \emph{stable heteroclinic cycle} if there is a closed attracting set $\Gamma_\text{het}$ composed of trajectories connecting a repeating sequence of saddle equilibrium points \cite{HolmesStone1992heteroclinic,KrupaMelbourne1995asymptotic,RabinovichHuertaVaronaAfraimovich2006BICY,ShawParkChielThomas2012SIADS,HorchlerDaltorioChielQuinn2015BioinspBiomim}. 
In this situation, there is no periodic trajectory with a finite period.  Instead, trajectories near $\Gamma_\text{het}$ traverse the same neighborhood of phase space with progressively longer and longer intervals required to pass each saddle point in turn.  Because there is no finite period, the phase and the asymptotic phase cannot be defined; see \cite{ShawParkChielThomas2012SIADS} for a discussion of the phase reduction problem for the deterministic case, and \cite{ThoLin14} for the stochastic case.  

A \emph{spiral sink system} possesses a stable equilibrium point for which the Jacobian matrix has a complex conjugate pair of eigenvalues.  As a simple example, consider the rotationally symmetric system 
\be
\label{eq:spiral_sink}
\frac{d\mbx}{dt}=\matrix{cc}{\mu & -\omega \\ \omega & \mu}\mbx,\quad \text{or,}\quad\dot{r}=\mu r,\quad\dot{\vartheta}=\omega
\ee
(in standard polar coordinates) with $\mu<0$ and $\omega>0$.  
It is well known that one cannot assign an asymptotic phase to points in the basin of attraction of a spiral sink fixed point such as \e{spiral_sink}, because (unlike for a limit cycle, cf.~\e{asymptotic_phase}) all initial points converge to the \emph{same} trajectory at long times, namely $\mbx(t)\to \mathbf{0}$ as $t\to \infty$.
If we seek solutions of \e{dTdt} for the vector field \e{spiral_sink}, we find a one-parameter family of solutions
\be
\label{eq:solutions_to_spiral_sink_phase}
\mathcal{T}(r,\vartheta)=k\vartheta + \frac{1-k\omega}{|\mu|}\ln r,
\ee
for arbitrary constant $k\in\R$.  The corresponding ``isochrons'' are given, in polar coordinates, as logarithmic spirals 
$\vartheta(r) = \vartheta_0+\frac{k\omega-1}{|\mu|}\ln r$.  Setting $k=1/\omega$ gives evenly spaced ``spokes of a wheel" isochrons, $\vartheta=\text{const}$, but this is only one choice in an infinite collection of solutions consistent with constant ``phase" progression.  

In contrast, we will show in the remainder of the paper that when a system with spiral sink dynamics is perturbed by noise, the isochrons of the forward and backward phase may be defined uniquely and unambiguously.  This surprising result is a major contribution of our paper.

\section{The model: two-dimensional spiral sink with white noise}
\label{sec:model}
We consider a two-dimensional Ornstein-Uhlenbeck process in a setting such that the origin becomes a stable sink. This is of course a classical stochastic process that has been well studied \cite{UhlOrn30}; much of the information we review in the following and adapt to our special notation can be found in standard text books, e.g.~\cite{Ris84} and \cite{Gar85}. 

The general Langevin equation is
\be
\dot{\vec{x}}=A \vec{x}+B \vec{\xi}, \;\;\;\;\; \lr{\xi_i(t)\xi_j(t')}=\delta_{i,j}\delta(t-t'), \;\; i,j=1,2.
\label{eq:OUP_2d}
\ee
We first discuss suitable choices and the properties of the two matrices $A$ and $B$.

We assume that the two eigenvalues of $A$ are complex
(they are then a complex conjugate pair). 
As we will show below, for the stochastic spiral sink, the eigenvalues of $A$ coincide with the slowest decaying eigenvalues of the forward and backward operators, discussed in \sse{stochastic_setting}.
Therefore, to avoid introducing superfluous notation, we will write the eigenvalues of $A$ as  $\lambda_\pm=\mu\pm i\omega$; recall that $\mu<0$ and $\omega>0$.
We will use a specific notation for the matrix $A$ by which we can write  $\lambda_\pm$ in a convenient way:
\ba
\label{eq:A_and_lambda}
A&=&\matrix{cc}{\mu+\alpha_\mu & -\omega_0+\alpha_\omega\\ \omega_0+\alpha_\omega&\mu-\alpha_\mu}, \\
 \lambda_\pm&=&\frac{\mathrm{Tr}(A)\pm i\sqrt{4\det(A)-(\mathrm{Tr}(A))^2}}{2}\nonumber\\&=&\mu\pm i \sqrt{\omega_0^2-(\alpha_\mu^2+\alpha_\omega^2)}=\mu\pm i\omega,
\ea
where the last equation defines the frequency of rotation $\omega$. Throughout, we require $\mu<0$ (stability of the fixed point at the origin) and $\omega_0>0$ (counter-clockwise rotation).  Setting $\alpha_\mu=\alpha_\omega=0$, we would obtain a rotationally symmetric stable sink, thus, $\{\alpha_\mu, \alpha_\omega\}$ quantify the deviation from this special case. Obviously, to keep complex eigenvalues, we have to require that $\omega_0^2>\alpha_\mu^2+\alpha_\omega^2$ which we will assume to hold true throughout the following.
We note that the phases will not change when we use a different unit of time and, hence,  we could set either the real or the imaginary part of the eigenvalue to unity, without loss of generality. However, for the  sake of broader applicability, we keep $\mu$ and $\omega_0$ in the following. 

The complex-valued left and right eigenvectors of $A$ satisfy 
\be
A\vec{v}_\pm=\lambda_\pm \vec{v}_\pm,\;\;\vec{w}_\pm^* A=\lambda_\pm \vec{w}_\pm^*,
\ee
(the asterisk denotes complex-conjugate transpose of a vector).  They are given in terms of our parameters by 
\be
\label{eq:ev}
\vec{v}_\pm=N e^{i\eta}\matrix{c}{\alpha_\omega-\omega_0\\ \pm i\omega-\alpha_\mu},\;\;\vec{w}_\pm=\matrix{c}{\alpha_\omega+\omega_0\\ \mp i\omega-\alpha_\mu}, 
\ee
where the normalization factors $N>0$ and $e^{i\eta}$ are chosen so that 
$\vec{w}_\pm^* \vec{v}_\pm = 1$, from which follows
$\eta=\pi-\mbox{atan}\left(\frac{\alpha_\mu}{\omega}\right)$.

Turning to the noise matrix $B$, we note that what enters the theory is only the symmetric diffusion matrix 
\ba
D&=&\frac12 B B^T\nonumber\\
&=&\frac12 \matrix{cc}{B_{11}^2+B_{12}^2 & B_{11}B_{21}+B_{12}B_{22}\\ B_{11}B_{21}+B_{12}B_{22} &B_{22}^2+B_{21}^2}\nonumber\\
&=&\varepsilon\matrix{cc}{1+\beta_D & \beta_c\\ \beta_c & 1-\beta_D}.
\ea 
Here we have expressed $D$ by deviations from the isotropic-noise case ($\beta_D=\beta_c=0$) that can occur when the noise is of different strength in the two variables (quantified by $-1\le\beta_D\le 1$) or when the noise in both variables is correlated ($\beta_c>0$) or anticorrelated ($\beta_c<0$), with the constraint $\beta_c^2+\beta_D^2\le 1$. We note that because the system is linear there is no qualitative change to be expected when we turn up the overall noise intensity $\varepsilon$. 
Without the fixed values of the decay rate $-\mu$ and noise intensity $\varepsilon$ we have the five parameters left: $\omega_0, \alpha_\mu,\alpha_\omega, \beta_D, $ and $\beta_c$. 
 
The  forward and backward  operators corresponding to \e{OUP_2d} read (using Einstein's summation convention)
\ba
\label{eq:fwd}
\LL[P(\mbx)]&=-\partial_i\left(A_{ij}x_jP(\mbx)\right)+D_{ij}\partial^2_{ij}P(\mbx)\\
\label{eq:bwd}
\LL^\dagger[Q^*(\mbx)]&=A_{ij}x_j\partial_i\left(Q^*(\mbx)\right)+D_{ij}\partial^2_{ij}Q^*(\mbx).
\ea
The covariance matrix $\Sigma_x$, corresponding to the stationary solution $\LL[P]=0$,
satisfies the Lyapunov equation 
\be
\label{eq:Lyapunov}
A\Sigma_x+\Sigma_xA^T+D=0.
\ee
Moreover, it can be expressed as  \cite{Gar85}
\ba
\Sigma_x&=&\lr{\vec{x}(t) \vec{x}^T(t)}\\ \nonumber
&=&-\frac{1}{\mathrm{Tr}(A)} D-
\frac{[A-\mathrm{Tr}(A) I]D[A-\mathrm{Tr}(A)I]^T}{ \mathrm{Tr}(A)\det(A)}.
\ea
In what follows, we will assume $\Sigma_x$ to be invertible.
 \begin{widetext} 
Inserting all the different terms, we arrive at
\ba
\Sigma_{x,1,1}&=&\frac{(1+\beta_D)[\mu(\mu-\alpha_\mu)+\alpha_\omega(\omega_0-\alpha_\omega)]+(\omega_0-\alpha_\omega)^2+\beta_c(\mu-\alpha_\mu)(\omega_0-\alpha_\omega)}{-\mu(\mu^2+\omega^2)}\nonumber\\
\Sigma_{x,2,2}&=&\frac{(1-\beta_D)[\mu(\mu+\alpha_\mu)-\alpha_\omega(\omega_0+\alpha_\omega)]+(\omega_0+\alpha_\omega)^2-\beta_c(\mu+\alpha_\mu)(\omega_0+\alpha_\omega)}{-\mu(\mu^2+\omega^2)}\nonumber\\
\Sigma_{x,1,2}=\Sigma_{x,2,1}&=&\frac{\alpha_\mu\omega_0-\alpha_\omega\mu-\beta_D(\mu\omega_0-\alpha_\omega\alpha_\mu)+\beta_c(\mu^2-\alpha_\mu^2)}{-\mu(\mu^2+\omega^2)}.
\label{eq:variance_x}
\ea
By means of the inverse of the covariance matrix
\ba
\Sigma_{x,1,1}^{-1}&=&-\mu\frac{(1-\beta_D)[\mu(\mu+\alpha_\mu)-\alpha_\omega(\omega_0+\alpha_\omega)]+(\omega_0+\alpha_\omega)^2-\beta_c(\mu+\alpha_\mu)(\omega_0+\alpha_\omega)}{(\omega_0+\alpha_\omega\beta_D-\alpha_\mu\beta_c)^2+\mu^2(1-\beta_c^2-\beta_D^2)}
\nonumber\\
\Sigma_{x,2,2}^{-1}&=&-\mu\frac{(1+\beta_D)[\mu(\mu-\alpha_\mu)+\alpha_\omega(\omega_0-\alpha_\omega)]+(\omega_0-\alpha_\omega)^2+\beta_c(\mu-\alpha_\mu)(\omega_0-\alpha_\omega)}{(\omega_0+\alpha_\omega\beta_D-\alpha_\mu\beta_c)^2+\mu^2(1-\beta_c^2-\beta_D^2)}\nonumber\\
\Sigma_{x,1,2}^{-1}=\Sigma_{x,2,1}^{-1}&=&\mu\frac{\alpha_\mu\omega_0-\alpha_\omega\mu-\beta_D(\mu\omega_0-\alpha_\omega\alpha_\mu)+\beta_c(\mu^2-\alpha_\mu^2)}{(\omega_0+\alpha_\omega\beta_D-\alpha_\mu\beta_c)^2+\mu^2(1-\beta_c^2-\beta_D^2)},
\label{eq:variance_inv_x}
\ea
 \end{widetext} 
we can express the stationary probability density as follows:
\be
P_0=\frac{\exp(-\frac12 \vec{x}^T \Sigma_x^{-1} \vec{x})}{2\pi \det(\Sigma_x)} .
\label{eq:P0}
\ee
We note that the case of a symmetric probability density is more general than the special case of the rotationally symmetric  drift matrix plus an isotropic noise. 
Formally, we may ask under what conditions the stationary variance will be a multiple of the identity, $\Sigma_x\propto I$.  Setting $\Sigma_{x,1,1}=\Sigma_{x,2,2}$ and $\Sigma_{x,1,2}=0$ in \e{variance_x} yields a unique solution
\be
\alpha_\mu=\mu\beta_D\quad\quad\text{and}\quad\quad\alpha_\omega=\mu\beta_c.
\label{eq:cond_isotrop_P0}
\ee
Thus a necessary condition for an isotropic distribution is that any asymmetry in the individual dissipation rates is compensated by a matched asymmetry in the individual noise terms ($\alpha_\mu=\mu\beta_D$), and any asymmetry in the coupling between the variables is compensated by an anticorrelation in the driving noise ($\alpha_\omega=\mu\beta_c$).

In the absence of a driving noise, the stationary distribution collapses to a delta distribution at the origin, and the covariance matrix reverts to the zero matrix.  As shown in 
 \sse{deterministic_setting}, in this case one can no longer uniquely define an asymptotic phase function.  However, as we will show in \se{analytic}, as long as the noise has finite amplitude and $\Sigma_x$ has full rank, the asymptotic phase obtained from the backward equation, as well as the phase from the forward equation, is well defined. 

The cross-spectral matrix with the power spectra of $x_1(t)$ and $x_2(t)$ on the diagonal can be obtained from the well-known expression \cite{Gar85}:
\be
S_{\textbf{xx}}(f)=2[A+2\pi i f I]^{-1} D [A^T-2\pi i f I]^{-1}.
\label{eq:power_spectrum}
\ee
Specifically, for the first element of the matrix, the power spectrum of the first variable, we obtain explicitly in terms of our parameters
\begin{widetext}
\be
S_{x_1x_1}(f)=2\frac{(\mu+\alpha_\mu)^2+(\omega_0-\alpha_\omega)^2+(2\pi f)^2+\beta_D((\mu+\alpha_\mu)^2-(\omega_0-\alpha_\omega)^2+(2 \pi f)^2)+2\beta_c(\alpha_\omega-\omega_0)(\mu+\alpha_\mu)
}{[\mu^2+\omega^2-(2\pi f)^2]^2+4\mu^2\omega^2}.
\label{eq:power_spectrum}
\ee
\end{widetext}
We note that the quality factor (the ratio between the peak frequency $f_{\textrm{peak}}$ and the full width at half maximum, $\Delta f$) of a robustly oscillatory OU process is very close to half the ratio of the imaginary and real parts of the system's eigenvalue, i.e. 
\be
Q_\text{F}=\frac{f_{\textrm{peak}}}{\Delta f}\approx \frac{\omega}{2\mu}.
\ee

\section{Analytical expressions for the asymptotic phases}
\label{sec:analytic}

We aim to find the eigenfunctions with the smallest negative real part for the forward and backward operators. In particular, we seek  the complex arguments of these functions, which define our forward and backward phases.
For the general system \e{OUP_2d}, no expresssions of the eigenfunctions in terms of elementary functions is known. However, Leen et al. \cite{LeeFri16}  recently derived   expressions for a simpler system that we will use. This system is given by
\be
\dot{\vec{y}}=A_y \vec{y}+B_y \vec{\xi}, \;\;\;\;\; \lr{\xi_i(t)\xi_j(t')}=\delta_{i,j}\delta(t-t'), \;\; i,j=1,2
\label{eq:OUP_2d_y}
\ee
and is simpler because its covariance matrix is a multiple of the identity matrix:
\be
\Sigma_y=\lr{\vec{y}(t) \vec{y}^T(t)}=\frac12 I.
\label{eq:sigma_y}
\ee 
We can transform our original model, \e{OUP_2d}, to \e{OUP_2d_y} obeying \e{sigma_y} as follows:
\be
y=Cx\;\; \mbox{with}\;\; C=\begin{pmatrix}
  \;  \vec{u}^T_{1}/\sqrt{2\gamma_{1}}\;\\[1em]
 \hline  \\
  \;  \vec{u}^T_{2}/\sqrt{2\gamma_{2}}\;\\
\end{pmatrix}
\ee
 where $\vec{u}_{k}$ and  $\gamma_k$ with $k=1,2$ are the  normalized eigenvectors and eigenvalues of the original (symmetric) covariance matrix $\Sigma_x$, respectively. Indeed $\Sigma_y=\lr{\vec{y}\vec{y}^T}=\lr{C\vec{x}\vec{x}^TC^T}=C\Sigma_x C^T$ and using the property $u^T_{i}u_{j}=\delta_{ij}$, we obtain \e{sigma_y}. Furthermore, 
with this transformation follow relations between drift and diffusion matrices of the original and the transformed system:
\be
A_y=CAC^{-1}, \;\; B_y=C B.
\ee
If, for long times, the conditional probability density in $\vec{y}$ is given by
\begin{align}
&P(\vec{y},t|\vec{y}_0,0) 
\simeq \\
&\quad P_0(\vec{y})+e^{\bar{\lambda}_+t}P_+(\vec{y}) Q^*_+(\vec{y}_0)+e^{\bar{\lambda}_-t}P_-(\vec{y}) Q^*_-(\vec{y}_0)  \nonumber
\end{align}
 where 
 $P_\pm(\vec{y})$ and  $Q_\pm(\vec{y}_0)$ are the corresponding eigenfunctions of forward and backward operators with smallest real part of the corresponding eigenvalues $\bar{\lambda}_\pm$. Applying eqs. (7), (9), and (12) in Leen et al. \cite{LeeFri16}, we obtain these eigenfunctions in terms of $\vec{v}_{y,\pm}=C\vec{v}_{\pm}$ and $\vec{w}^*_{y,\pm}=\vec{w}^*_{\pm}C^{-1}$, the corresponding right and left eigenvectors of the transformed drift matrix $A_y=CAC^{-1}$. They read:
\ba
Q_0(\vec{y}_0)&=&1,\\
P_0(\vec{y})&=&\frac{e^{-\vec{y}^T\vec{y}}}{\pi},\\
P_\pm(\vec{y})&=&-\frac{2}{\pi}e^{-\vec{y}^T\vec{y}}\vec{y}^T\vec{v}_{y,\pm} ,\\ Q^*_\pm(\vec{y}_0)&=&2\vec{w}^{*}_{y,\pm} \vec{y}_0.
\ea
Inserting these expressions into \e{fwd}-\e{bwd}, together with \e{Lyapunov} and the fact that $\Sigma_y=\frac12I$, it can be checked that these functions satisfy the forward and backward equations, respectively, with eigenvalues $\lambda_0=0$ (for $Q_0$ and $P_0$) and $\bar{\lambda}_\pm=\lambda_\pm$ (for $P_\pm$ and $Q^*_\pm$, respectively) identical to the eigenvalues of $A$ given in \e{A_and_lambda}.  

From the above asymptotic expansion of the density $P(\vec{y},t|\vec{y}_0,0)$ and its   transformation $P(\vec{x},t|\vec{x}_0,0)=P(\vec{y}(\vec{x}),t|\vec{y}_0(\vec{x}_0),0)|d\vec{y}/d\vec{x}|$, we obtain
\ba
&&P(\vec{x},t|\vec{x}_0,0)\approx \frac{1}{\pi} \left|\frac{d\vec{y}}{d\vec{x}}\right|\exp\left[-\frac12\vec{x}^T\Sigma_x^{-1}\vec{x}\right]\nonumber\\
&&\times\left(1-4\left\{e^{\lambda_+t}(\vec{x}^TC^T\vec{v}_{y,+})(\vec{w}_{y,+}^{*}C\vec{x}_0)+c.c.\right\}\right).
\ea
Noting that for a linear transformation the Jacobian $|d\vec{y}/d\vec{x}|=const$, we can conclude that the first eigenfunctions in the original variables read 
\ba
P_+(\vec{x})&\propto&  
\vec{x}^TC^T\vec{v}_{y,+}e^{-\frac12\vec{x}^T\Sigma_x^{-1}\vec{x}} 
= \vec{x}^T\Sigma_x^{-1}\vec{v}_{+}e^{-\frac12\vec{x}^T\Sigma_x^{-1}\vec{x}},\nonumber\\
Q^*_+(\vec{x}_0)&\propto& 
\vec{w}^*_{y,+}C\vec{x}_0=\vec{w}^*_{+}\vec{x}_0.
\ea

The forward and backward phases can now be extracted from these expressions taking into account that the only complex-valued objects in them are   $\vec{v}_+$ and $\vec{w}^*_+$. Specifically, for the backward phase we obtain the concise expression in terms of the geometric phase $\vartheta$:
\begin{widetext}
\be
\Psi(\vartheta)=\mbox{atan}\left(\frac{\omega}{(\alpha_\omega+\omega_0)\cot(\vartheta)-\alpha_\mu}\right)
\label{eq:BW_phase}
\ee
while the expression for the forward phase, involving the inverse of the covariance matrix, is more lengthy
\be
\label{eq:FW_phase}
\Phi(\vartheta)=\pi-\mbox{atan}\left(\frac{\alpha_\mu}{\omega}\right)-\mbox{atan}\left(\frac{A_1+\tan(\vartheta)}{A_2+A_3\tan(\vartheta)}\right)
\ee
where
\ba
&&A_1=\frac{\beta_c(\alpha_\mu^2-\mu^2)+\beta_D(\mu\omega_0-\alpha_\mu\alpha_\omega)-\alpha_\mu\omega_0+\alpha_\omega\mu}{\beta_D(\alpha_\omega(\omega_0-\alpha_\omega)+\mu(\mu-\alpha_\mu))+\beta_c(\mu-\alpha_\mu)(\omega_0-\alpha_\omega)+\mu(\mu-\alpha_\mu)+\omega_0(\omega_0-\alpha_\omega)},\nonumber\\
&&A_2=\frac{1}{\omega}\frac{\mu\left[\beta_D(\mu\omega_0-\alpha_\mu\alpha_\omega)+\beta_c(\omega_0^2-\alpha_\omega^2)-\alpha_\mu\omega_0+\alpha_\omega\mu\right]+(\omega^2+\mu^2)(\beta_c\alpha_\mu-\beta_D\alpha_\omega-\omega_0)}{\beta_D(\alpha_\omega(\omega_0-\alpha_\omega)+\mu(\mu-\alpha_\mu))+\beta_c(\mu-\alpha_\mu)(\omega_0-\alpha_\omega)+\mu(\mu-\alpha_\mu)+\omega_0(\omega_0-\alpha_\omega)},\nonumber\\
&&A_3=-\frac{\mu}{\omega}\frac{\beta_D(\omega_0(\omega_0-\alpha_\omega)+\alpha_\mu(\mu-\alpha_\mu))-\beta_c(\omega_0-\alpha_\omega)(\mu-\alpha_\mu)+\alpha_\mu(\mu-\alpha_\mu)+\alpha_\omega(\omega_0-\alpha_\omega)}{\beta_D(\alpha_\omega(\omega_0-\alpha_\omega)+\mu(\mu-\alpha_\mu))+\beta_c(\mu-\alpha_\mu)(\omega_0-\alpha_\omega)+\mu(\mu-\alpha_\mu)+\omega_0(\omega_0-\alpha_\omega)}.\nonumber
\ea
It will prove useful to have the expressions for the derivatives, that can be interpreted as densities of the respective isochrons. For the backward phase, we obtain
\be
\frac{d\Psi}{d\vartheta}=\frac{\omega}{\omega_0-\alpha_\omega+2\alpha_\omega\cos^2(\vartheta)-2\alpha_\mu\cos(\vartheta)\sin(\vartheta)},
\label{eq:BW_deriv}
\ee
whereas for the forward phase, this derivative reads
\be
\frac{d\Phi}{d\vartheta}=\frac{A_1A_3-A_2}{1+A_3^2+[A_1^2+A_2^2-A_3^2-1]\cos^2(\vartheta)+2[A_2A_3+A_1]\sin(\vartheta)\cos(\vartheta)}.
\label{eq:FW_deriv}
\ee

\end{widetext}

A number of conclusions can be drawn analytically from eqs.~(\ref{eq:BW_phase}-\ref{eq:FW_deriv}):
\begin{enumerate}[leftmargin=*,labelindent=0pt,itemsep=0pt]
\item The asymptotic phase of a stochastic oscillator is uniquely defined in terms of the backward phase \e{BW_phase}. Hence, in marked contrast to the deterministic case, 
where we obtain an entire family of possible phase definitions (cf. \e{solutions_to_spiral_sink_phase} and the surrounding discussion), we do not have any ambiguity in the phase definition in the stochastic case, except for an arbitrary additive off-set. As we will see, our asymptotic phase defines the phase uniquely even in the case of vanishing noise and extracts the most simple, i.e. the non-logarithmic, definition of phase (see points 2 and 3 below).    
\item The isochrons (the lines of equal phase) are rays starting from the origin: a rescaling of the vector $\vec{x}_0$ (which describes a straight line starting in the origin), for instance, cannot change the complex phase of the complex number $\vec{w}_+^*\vec{x}_0$ (this factor just rescales the complex number itself); the same argument holds true for the forward phase, i.e.~the complex argument of $\vec{x}^T\Sigma_x^{-1}\vec{v}_+$ upon varying $\vec{x}$. Another manifestation of this simple geometry is that both forward and backward phases are only functions of the geometric phase but not of the radial coordinate. 
\item  Changing the geometric phase $\vartheta$ of the vector $\vec{x}_0$, however, we do not necessarily get a proportional change in the complex argument of  $Q^*_+$. This means that although the isochrons are the spokes of a wheel, these spokes are sometimes closer together and sometimes further apart, i.e. the functions $\Psi(\vartheta)$ and $\Phi(\vartheta)$ are only in exceptional cases linear (the phase densities $d\Psi(\vartheta)/d\vartheta$ and $d\Phi(\vartheta)/d\vartheta$ are then constant).
\item The backward phase cannot depend on the noise properties of the system, because it is entirely determined by the drift matrix $A$, more specifically, by the left eigenvector of $A$. Furthermore, the backward phase is also independent of the decay rate $-\mu$ (which does not enter the eigenvectors) but depends only on $\omega_0, \alpha_\omega,$ and $\alpha_\mu$.
\item The geometric phase will not progress at a steady pace, even in the deterministic case. To see this, transform the system to polar coordinates to find in the deterministic case ($\varepsilon=0$)
\be
\frac{d\vartheta}{dt}=\omega+\alpha_\omega[\cos^2(\vartheta)-\sin^2(\vartheta)]-2\alpha_\mu\sin(\vartheta)\cos(\vartheta).
\nonumber
\ee 
Because of the nonlinearities on the right hand side, progress in the geometric phase will not be constant and this will not be different in general for a finite noise intensity $\varepsilon>0$. In contrast, taking the time derivative of the complex-valued backward function, that we write as a product of an amplitude $u$ and the complex exponential of the phase $\Psi$,  we obtain
\ba
\frac{dQ^*_+}{dt}&=&\frac{d}{dt} u e^{i\Psi(t)}=\underline{\frac{du}{dt}e^{i\Psi(t)}+i\frac{d\Psi}{dt} u e^{i\Psi(t)}}\nonumber\\
&=&\frac{d}{dt} \vec{w}^*_+ \vec{x}\nonumber\\
&=&\vec{w}^*_+( A \vec{x}+B\vec{\xi})=\lambda_+\vec{w}^*_+ \vec{x} +\vec{w}^*_+ B\vec{\xi}\nonumber\\
&=&\underline{(\mu+i\omega) u e^{i\Psi(t)} +\vec{w}^*_+ B\vec{\xi}}.
\ea
Focusing on the underlined parts, dividing by the complex exponential, and averaging over the noise, we arrive at two simple equations for the time-dependent mean values
\be
\frac{d\lr{u}}{dt}=\mu \lr{u},\quad\quad \frac{d\lr{\Psi}}{dt}=\omega.
\ee
Hence, the mean amplitude decays exponentially and the average backward phase evolves at a constant velocity.
\item The forward phase \emph{does} in general  depend on the noise matrix $D$, as well as on the drift matrix $A$, because it involves the inverse of the covariance matrix that is shaped by the coefficients in the matrix $D$. Hence, in general forward and backward phases are distinct. 
\item With a few algebraic manipulations and a comparison of coefficients of trigonometric functions in the  derivatives \e{FW_deriv} and \e{BW_deriv}, it can be shown that for an isotropic noise ($\beta_D=\beta_c=0$), forward and backward  phases  (apart from their off-set) are shifted versions of each other with respect to the geometric phase, i.e. the derivatives are 
\be
\left.\frac{d\Psi}{d\vartheta}\right|_{\vartheta+\arctan(\mu/\omega_0)}=\left.\frac{d\Phi}{d\vartheta}\right|_{\vartheta}\quad \mbox{for}\quad \beta_D=\beta_c=0.
\label{eq:phase_shift}
\ee 
\item In the limit  as $\mu\to 0$ (hence, quality factor $Q_\text{F}\to\infty$) both forward and backward densities are equal 
\be
\frac{d\Psi(\vartheta)}{d\vartheta}=\frac{d\Phi(\vartheta)}{d\vartheta} \quad \mbox{for}\quad \mu=0.
\ee
If $\mu$ is small compared to the effective frequency $\omega$, i.e. for large quality factor, the two phases will be close to each other.
\item For the case of a rotationally symmetric probability density, i.e. when  \e{cond_isotrop_P0} holds true, it can be shown that the densities of the phase are shifted versions of each other and the shift is $\pi/2$:
 \be
\left.\frac{d\Psi}{d\vartheta}\right|_{\vartheta\pm \pi/2}=\left.\frac{d\Phi}{d\vartheta}\right|_{\vartheta}\quad \mbox{for}\quad \alpha_\mu=\mu\beta_D, \alpha_\omega=\mu\beta_c.
\label{eq:phase_shift2}
\ee 

\end{enumerate}

\section{Illustration of specific cases}
\label{sec:illustration}
Here we give several numerical examples of isochrons, their density, and the stationary probability density of the system. It is useful to classify the distinct cases of anisotropy that can be caused by anisotropic terms in the drift matrix ($\alpha_\mu,\alpha_\omega$) or in the diffusion matrix ($\beta_D, \beta_c$) by the two parameter vectors 
\be
\vec{\alpha}=\matrix{c}{\alpha_\mu\\ \alpha_\omega},\quad \vec{\beta}=\matrix{c}{\beta_D\\ \beta_c}.
\ee
In the following, we demonstrate that i) for the complete isotropic case ($\vec{\alpha}=0, \vec{\beta}=0$) both phase isochrons are uniformly distributed (as can be expected); ii) for a uniform  density of isochrons of the backward phase, the forward phase's isochrons can be  non-uniformly distributed (e.g. for ($\vec{\alpha}=0, \vec{\beta}\neq 0$); iii) for isotropic noise ($\vec{\beta}=0$) but anisotropic drift matrix ($\vec{\alpha}\neq 0$) both isochron densities are non-uniform, but are shifted versions of each other; iv) for an  isotropic stationary distribution $P_0$ (which happens for $\vec{\alpha}=\mu\vec{\beta}$) there can be non-uniform densities of backward and forward phases; v) backward and forward phase are particulary different for $\vec{\alpha}=-\mu\vec{\beta}$.

\begin{figure*} 
   \centering
  \includegraphics[width=1\textwidth]{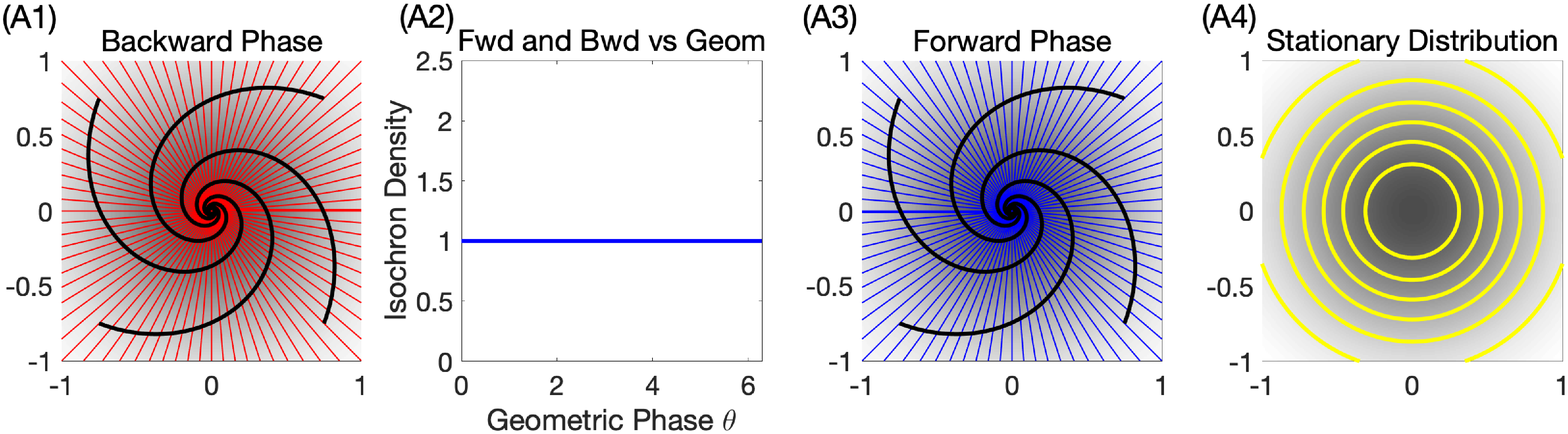} 
   \includegraphics[width=1\textwidth]{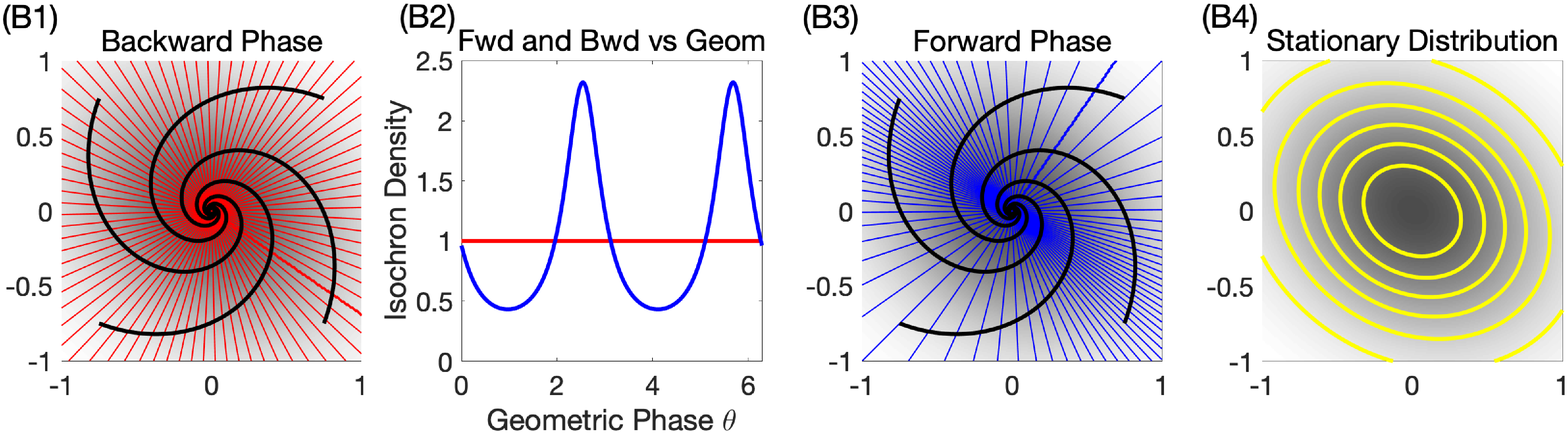} 
   \includegraphics[width=1\textwidth]{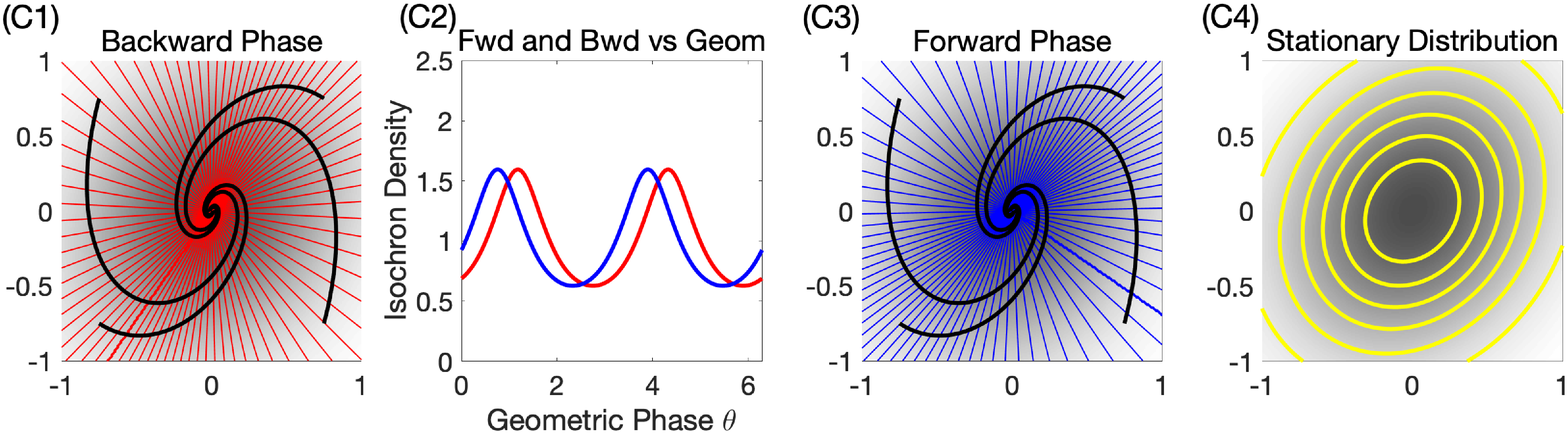} 
   \caption{\textbf{Isotropic drift and diffusion matrices (first row, A), isotropic drift and anisotropic diffusion (second row, B), anisotropic drift and isotropic diffusion (third row, C)}  Shown are the isochrons for the backward phase (first column, A1-C1) and the forward phase (third column, A3-C3), the densities of the isochrons $d\Psi/d\vartheta$ and $d\Phi/d\vartheta$ as functions of the geometric phase (second column, A2-C2) according to \e{BW_deriv} and \e{FW_deriv}, respectively, and the stationary probability densities $P_0(x_1,x_2)$ according to \e{P0} (fourth column, A4-C4) with contour lines in yellow; $P_0(x_1,x_2)$ (grey scale) and typical mean trajectories (black lines) also shown in  first and third column for comparison.   
Parameters, first row: $\alpha_\mu=\alpha_\omega=\beta_D=\beta_c=0$; second row: 
$\alpha_\mu=\alpha_\omega=0, \beta_D=\beta_c=-0.68493$; third row: $\alpha_\mu=\alpha_\omega=0.30822, \beta_D=\beta_c=0$;  in all panels $\omega_0=1, \mu=-0.45$. Horizontal and vertical axes in first, third and fourth rows are $x_1$ and $x_2$, respectively; points on all axes are given in arbitrary units.   
}
   \label{fig:example_1_3}
\end{figure*}

Starting with the first case of a completely isotropic system ($\vec{\alpha}=\vec{\beta}=0$), we show in the first row of \bi{example_1_3} the isochrons of the backward (A1) and forward (A3) phases in the $(x_1,x_2)$ plane, their density (i.e. their derivative with respect to the geometric phase $\vartheta$) as a function of $\vartheta$ (A2), and the stationary probability density (A4) vs. $(x_1,x_2)$. In this case, not surprisingly, we find a rotationally symmetric probability density and uniformly spaced spokes of a wheel as the isochrons of both forward and backward phase; the densities for both phases are flat. This is a direct demonstration that our asymptotic phase singles out the simplest of the possible solutions from the family of functions, \e{solutions_to_spiral_sink_phase}, possible for the deterministic system.

For an isotropic drift matrix ($\vec{\alpha}=0$) but anisotropic diffusion matrix ($\vec{\beta}\neq 0$), the asymptotic phase (backward phase) will remain  uniform because it does not depend on $\beta$ (cf. conclusion 4 in \se{analytic}) and this we can indeed see in \bi{example_1_3}B1 and B2.  The forward phase, in contrast, depends on $\beta$ and is now nonuniform as becomes apparent in \bi{example_1_3}B2 and B3. Also with an anisotropic noise we obtain in general an anisotropic probability density (cf. \bi{example_1_3}B4) as can be expected due to the correlations and unequal intensities of the noise sources driving the two components of the Ornstein-Uhlenbeck process.
 \begin{figure*}
   \centering
  \includegraphics[width=1\textwidth]{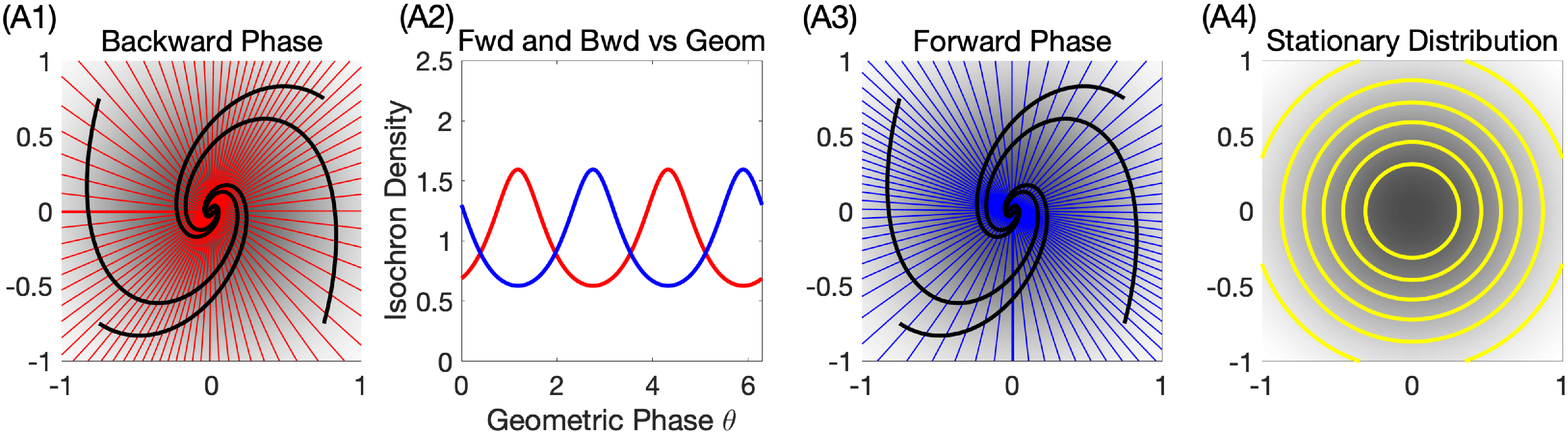} 
   \includegraphics[width=1\textwidth]{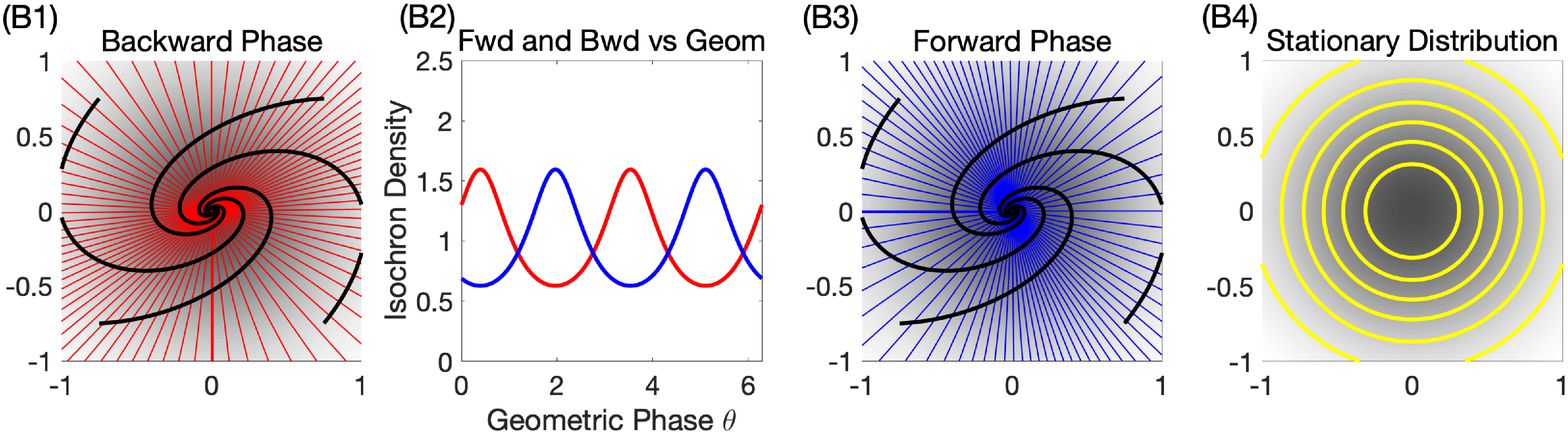} 
   \caption{\textbf{Non-uniform densities of backward and forward phases coexist with an isotropic stationary probability density of the system.} Two different cases of $\vec{\alpha}=\mu\vec{\beta}$ where we rotated the vectors used in (A1-A4) by $\pi/2$ for the panels (B1-B4). Shown are the isochrons for the backward phase (A1,B1) and the forward phase (A3,B3), the densities of the isochrons $d\Psi/d\vartheta$ and $d\Phi/d\vartheta$ (A2,B2) as functions of the geometric phase according to \e{BW_deriv} and \e{FW_deriv}, respectively, and the stationary probability densities $P_0(x_1,x_2)$ according to \e{P0}  (A4,B4); $P_0(x_1,x_2)$ (grey scale) and typical mean trajectories (black lines) also shown in  first and third column for comparison.  
Parameters in all panels: $\omega_0=1, \mu=-0.45$. First row: $\beta_D=\beta_c=-0.68493, \alpha_\mu=\alpha_\omega=\mu \beta_D$.  Second row: $ \beta_c=-\beta_D=0.68493, \alpha_\mu=-\alpha_\omega=\mu \beta_D$. Horizontal and vertical axes in first, third and fourth rows are $x_1$ and $x_2$, respectively; points on all axes are given in arbitrary units.}
   \label{fig:example_isotrop_P0}
\end{figure*}

Next, we choose an isotropic diffusion matrix ($\vec{\beta}=0$) but anisotropic drift matrix ($\vec{\alpha}\neq 0$). In this case, in accordance with conclusion 7 in \se{analytic}, \e{phase_shift}, both phase densities are shifted versions of each other as becomes evident in  \bi{example_1_3}C1-3. The probability density is again anisotropic  (\bi{example_1_3}C4).  

As we have observed in \se{model}, for a specific relation between drift and diffusion anisotropies, \e{cond_isotrop_P0}, or in terms of our vector notation, for
\be
\vec{\alpha}=\mu\vec{\beta},
\label{eq:cond_isotrop_P0_2}
\ee
the stationary probability density is isotropic despite the possible anisotropies of the single $A$ and $D$ matrices. Indeed, if we combine the two anisotropies from \bi{example_1_3}B and C, we obtain parameters that obey this condition, and consequently, the stationary probability density is rotationally symmetric (\bi{example_isotrop_P0}A4). In this case, the isochrons of the two phases are still nonuniformly distributed (\bi{example_isotrop_P0}A2) and the directions of maximum density are orthogonal to each other (cf. \bi{example_isotrop_P0}A1 and A3). Indeed, as we have demonstrated in conclusion 9 in \se{analytic}, the two isochron densities are identical after a shift by $\pm \pi/2$ in the argument $\vartheta$, \e{phase_shift2}.    

For comparison, \bi{example_isotrop_P0}B shows what happens if we rotate the $\alpha$ and $\beta$ vectors by $\pi/2$ (with $\alpha$ and $\beta$ still obeying the condition \e{cond_isotrop_P0_2}): evidently, the probability density is still isotropic (\bi{example_isotrop_P0}B4) while the isochrons have been rotated by $\pi/4$  (see \bi{example_isotrop_P0}B1 and B3) and the isochron densities have merely shifted by the  same angle. 

We finally look at a case that seems to be far away from matching \e{cond_isotrop_P0_2} by switching the sign of $\vec{\beta}$ compared to \bi{example_isotrop_P0}. Then we get a pronounced anisotropy in the stationary probability density (cf. \bi{example_anisotrop_P0}A4) and the densities of the isochrons of backward and forward phases look now qualitatively different (\bi{example_anisotrop_P0}A2). These densities are not merely shifted versions of each other,  but the forward phase displays much stronger variations in its density, which also can be seen on comparison of the spacing of the isochrons shown in \bi{example_anisotrop_P0}A1 and A3. As before, an equal rotation of the vectors $\vec{\alpha}$ and $\vec{\beta}$ (as done in (\bi{example_anisotrop_P0}B) does not change the picture qualitatively, but everything
(isochrons, stationary density) is rotated. 



\begin{figure*}
   \centering
  \includegraphics[width=1\textwidth]{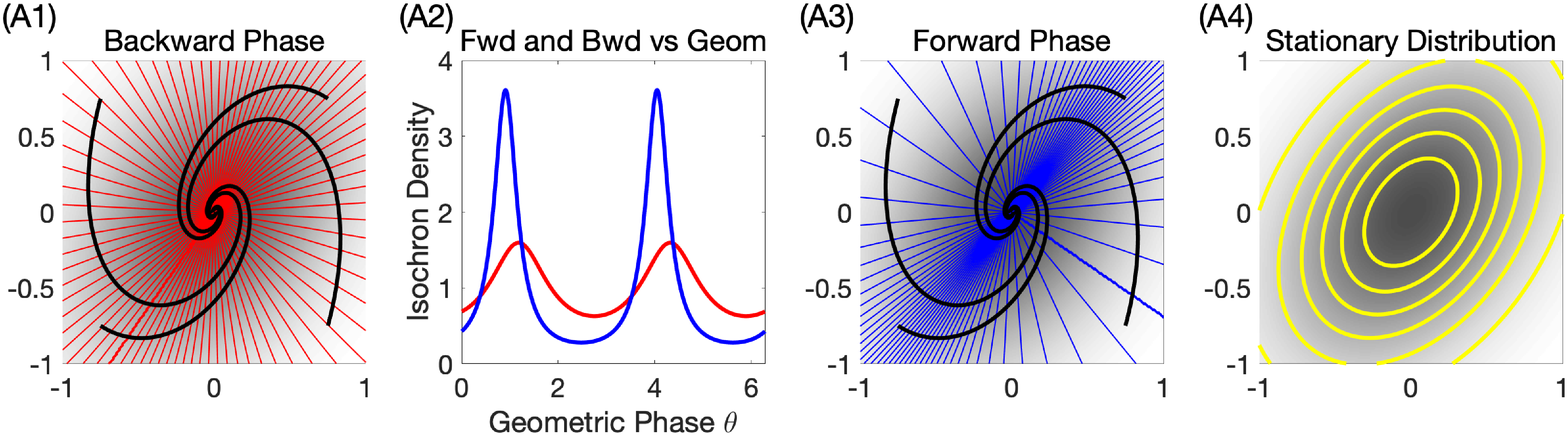} 
   \includegraphics[width=1\textwidth]{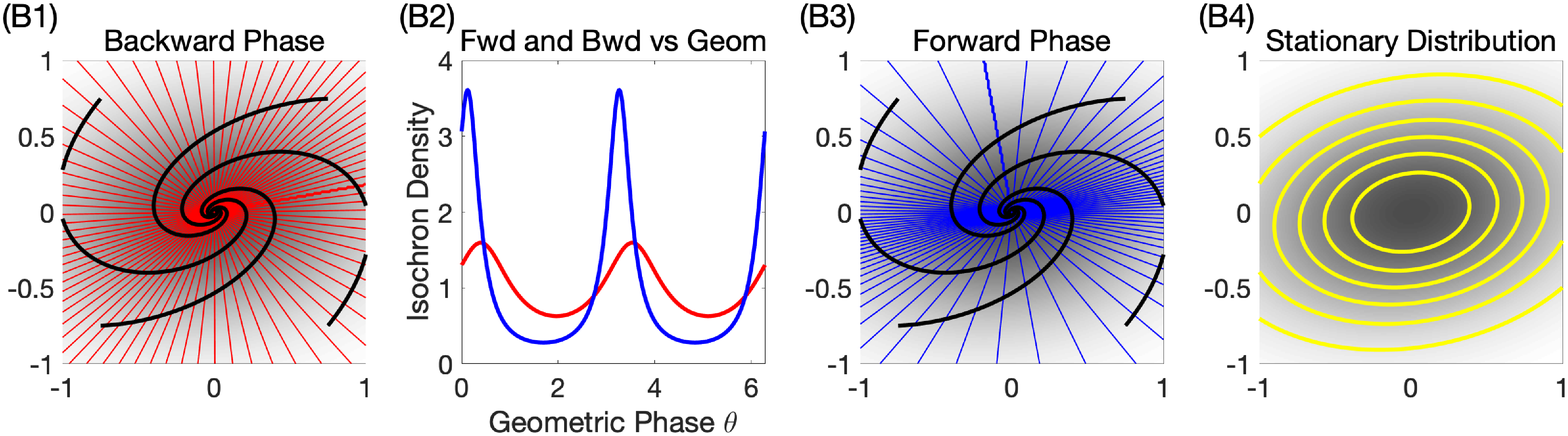} 
   \caption{\textbf{Reversing the relation between the anisotropy vectors leads to enhanced anisotropy in the system.} Two different cases of $\vec{\alpha}=-\mu\vec{\beta}$ (vector $\beta$ is switched in sign compared to \bi{example_isotrop_P0}). As before  we rotated the vectors used in (A1-A4) by $\pi/2$ for the panels (B1-B4). Shown are the isochrons for the backward phase (A1,B1) and the forward phase (A3,B3), the densities of the isochrons $d\Psi/d\vartheta$ and $d\Phi/d\vartheta$ (A2,B2) as functions of the geometric phase according to \e{BW_deriv} and \e{FW_deriv}, respectively, and the stationary probability densities $P_0(x_1,x_2)$ according to \e{P0}  (A4,B4); $P_0(x_1,x_2)$ (grey scale) and typical mean trajectories (black lines) also shown in  first and third column for comparison.  
Parameters in all panels: $\omega_0=1, \mu=-0.45$. First row: $\beta_D=\beta_c=0.68493, \alpha_\mu=\alpha_\omega=-\mu \beta_D$.  Second row: $ \beta_c=-\beta_D=0.68493, \alpha_\mu=-\alpha_\omega=-\mu \beta_D$. Horizontal and vertical axes in first, third and fourth rows are $x_1$ and $x_2$, respectively; points on all axes are given in arbitrary units.
  }
   \label{fig:example_anisotrop_P0}
\end{figure*}

\section{Conclusions}
\label{sec:conclusions}
In this paper we have derived and studied the asymptotic phase for a  simple stochastic oscillator, the two-dimensional Ornstein-Uhlenbeck process. Difficulties that are pertinent to the deterministic version of this system (the linearized version of a spiral sink) can be resolved by introducing our unambiguous asymptotic phase and taking the limit  of a vanishing noise intensity. Our phase, defined by the complex argument of the eigenfunction  of the backward Kolmogorov operator, singles out the simplest solution (spokes of a wheel) from the entire family of possible phases, \e{solutions_to_spiral_sink_phase}. To obtain the deterministic phase we do not have to let the noise intensity go to zero because (for the specific system considered here, the two-dimensional Ornstein-Uhlenbeck process) the asymptotic phase does not depend on the intensity and correlation properties of the driving white Gaussian noise at all. 

Although the asymptotic phase is independent of the radial variable leading to the spokes of a wheel for the isochrons, there is a nontrivial  dependence  of the density of isochrons as a function of the geometric phase angle. The density change is such that the mean progression in the asymptotic phase is always uniform - in contrast to the progression of the geometric phase  if the drift function deviates from the isotropic case (i.e. when $\alpha_\mu\neq 0$ or $\alpha_\omega\neq 0$). This can be regarded as the defining feature of the asymptotic phase. 

We also studied the phase related to the forward Kolmogorov operator's eigenfunction with smallest real part of the eigenvalue, or the forward phase, for short. The isochrons for this phase  turned out to be the spokes of a wheel as well, but the density of the spokes as a function of geometric phase was shown to be generally different from that of the backward phase and to be dependent on the diffusion matrix (i.e. on the intensities and correlations among the noise forces driving the two components of the system). This marked difference between forward and backward phase can be understood based on the role of forward and backward eigenfunctions in the evolution of the system, \e{P_asymp}. The forward eigenfunction which characterizes the evolution at long times forward in time is shaped by how correlated the noise may be in the two components. The backward phase (the true asymptotic phase) characterizes the initial conditions that brought the system to the considered point in time. In our examples we also demonstrated that even for a system with perfectly isotropic stationary probability density, the density of backward and forward phase may depend non-uniformly on the geometric phase.  

We identified several cases where the densities of backward and forward phases are merely shifted versions of each other: for a rotationally symmetric stationary probability density (with a shift of $\pi/2$), for an isotropic noise matrix (with a shift of $\arctan(\mu/\omega_0)$), and in the limit of infinite quality factor by $\mu\to 0$ (with a vanishing phase shift).  

A simple extension of the results considered here would be an $n$-dimensional Ornstein-Uhlenbeck process with $n>2$ when two of the eigenvalues of the Kolmogorov operator form a complex conjugate pair with real part that is much smaller in absolute value than the real parts of all other eigenvalues. In this case, the system would rapidly collapse to a two-dimensional manifold (the slow fiber, see \cite{Wiggins1994NormallyHyperbolicDS}), the phase reduction can be accomplished by projection onto this plane, and the isochrons of the asymptotic phase will be spokes of a wheel in this plane. 

Our results may also prove useful  in \emph{nonlinear}  stochastic systems  where a spiral sink may coexist with other dynamics, e.g.~with a limit cycle. For example, the subthreshold dynamics of the stochastic planar Morris-Lecar model can be described by a two-dimensional Ornstein-Uhlenbeck process \cite{DitlevsenGreenwood2013JMB}. 
Subthreshold oscillations have been  experimentally manipulated by
Stiefel et al.~\cite{StiefelFellousThomasSejnowski2010EJNsci} as a way of using  phase resetting (by inhibitory stimuli) as a mechanism for extended temporal integration. \emph{Phase resetting} presumes the existence of a \emph{phase} even though until now there was no well-defined notion of phase related to subthreshold oscillations (essentially a noise-driven dynamics close to a spiral sink). Our results provide a framework for further investigations in this direction.

\section{Acknowledgments} PJT was funded by NSF grant DMS-1413770.

\end{document}